\documentclass[%
 reprint,
showpacs,preprintnumbers,
 amsmath,amssymb,
 aps,
]{revtex4-1}

\usepackage{subfigure}
\usepackage[rflt]{floatflt}
\usepackage{color}

\usepackage{graphicx}
\usepackage{dcolumn}
\usepackage{bm}
\definecolor{darkblue}{rgb}{0.149019,0,0.439215}
\usepackage[dvipdfm,colorlinks=true,citecolor=darkblue,linkcolor=darkblue,urlcolor=darkblue]{hyperref}

\def\112{SrCuO$_2$}
\def\213{Sr$_2$CuO$_3$}

\begin{document}


\title{Bond disorder and spinon heat transport in the $S=\tfrac{1}{2}$ Heisenberg spin chain compound \213: from clean to dirty limits}

\date{\today}

\author{A. Mohan$^1$, N. Sekhar Beesetty$^2$, N. Hlubek$^1$, R. Saint-Martin$^2$, A. Revcolevschi$^2$, B. B\"{u}chner$^{1,3}$, C. Hess$^{1,3}$}

\address{$^1$Leibniz Institute for Solid State and Materials Research IFW Dresden, 01069 Dresden, Germany\\
$^2$Laboratoire de Physico-Chimie des Solides, Universit\'{e}
Paris-Sud, 91405 Orsay, Cedex, France\\
$^3$Center for Transport and Devices, Technische Universit\"{a}t
Dresden, 01069 Dresden, Germany}

\begin{abstract}

We investigate the effect of disorder on the heat transport
properties of the $S=\tfrac{1}{2}$ Heisenberg chain compound \213
upon chemically substituting Sr by increasing concentrations of
Ca. As Ca occupies sites outside but near the Cu-O-Cu spin chains,
bond disorder, i.e. a spatial variation of the exchange
interaction $J$, is expected to be realized in these chains. We
observe that the magnetic heat conductivity
($\kappa_{\mathrm{mag}}$) due to spinons propagating in the chains
is gradually but strongly suppressed with increasing amount of Ca,
where the doping dependence can be understood in terms of
increased scattering of spinons due to Ca-induced disorder. This
is also reflected in the spinon mean free path which can be
separated in a doping independent but temperature dependent
scattering length due to spinon-phonon scattering, and a
temperature independent but doping dependent spinon-defect
scattering length. The latter spans from very large ($>$ 1300
lattice spacings) to very short ($\sim$ 12 lattice spacings) and
scales with the average distance between two neighboring Ca atoms.
Thus, the Ca-induced disorder acts as an effective defect within
the spin chain, and the doping scheme allows to cover the whole
doping regime between the clean and the dirty limits.
Interestingly, at maximum impurity level we observe, in Ca-doped
\213, an almost linear increase of $\kappa_{\mathrm{mag}}$ at
temperatures above 100 K which reflects the intrinsic low
temperature behavior of heat transport in a Heisenberg spin chain.
These findings are quite different from that observed for the
Ca-doped double spin chain compound, \112, where the effect of Ca
seems to saturate already at intermediate doping levels.

\end{abstract}

\maketitle

\section{Introduction}

Quasi-particle transport in quantum systems has been a long
standing unsolved issue from the point of view of both theory and
experiment \cite{Sologubenko2000,Sologubenko2001,Hlubek2011,
Hlubek2010,Hlubek2012,Kawamata2008,Zotos1999,Alvarez2002,Azuma1998,
Castella1995,Chernyshev2005,Heidrich-Meisner2003,Heidrich-Meisner2002,
Hess2003,Hess2003a,Hess2001,Hess2007a,Hess2006,Louis2006,Klumper2002,
Orignac2003,Rozhkov2005,Saito2003,Saito1996,Sakai2003,Shimshoni2005,
Shimshoni2003,Zotos1997}. Much of the current interest is driven
by theoretical discoveries of anomalous transport properties in
quantum integrable systems \cite{Castella1995,Saito1996}.
Ballistic heat transport at finite temperatures in integrable spin
models, like the $S=\tfrac{1}{2}$ Heisenberg chain, is a
remarkable result that has received a lot of theoretical and
experimental attention. Due to the integrability of the
$S=\tfrac{1}{2}$ Heisenberg chain, the intrinsic spinon heat
transport is predicted to be non-dissipative, causing the heat
conductivity to diverge \cite{Zotos1999,Zotos1997}. Experimental
evidence of exceptionally large spinon heat conductivity and mean
free path in 1D spin chains of the highly pure cuprate materials
\112 and \213, has further strengthened this claim
\cite{Hlubek2010,Hlubek2012,Kawamata2008,Sologubenko2000,
Sologubenko2001}. Though in a real system ever present extrinsic
scattering mechanisms involving phonons and defects render the
system non-integrable, thereby limiting the heat conductivity to
finite values.

The large spinon heat conductivity in these systems also gives us
a chance to systematically investigate extrinsic scattering
processes, involving phonons and disorder, that spinons undergo to
relax the heat current. Experimental investigations on these
compounds by doping impurities in a controlled fashion have shown
that a quantitatively reasonable description of the temperature
dependence of spinon mean free paths is possible by taking into
account the scattering of magnetic excitations off phonons and
impurities alone \cite{Hess2007a,Hlubek2012,Kawamata2008,
Sologubenko2000,Sologubenko2001}. Such quantitative descriptions
falling under the framework of a semi-classical linearized
Boltzmann theory, in spite of being empirical in nature, are able
to model experimental observations with remarkable accuracy, thus
becoming a valuable input to the theory of quasiparticle heat
transport. A microscopic and fully quantum mechanical
understanding of the underlying scattering processes in quasi
one-dimensional systems is incomplete though. There do exist
theoretical studies on transport of spinons in $S=\tfrac{1}{2}$
Heisenberg chains coupled to phonons and impurities, which support
the experimental observation that scattering by impurities at very
low temperatures and by phonons at higher temperatures are the
processes that dictate the temperature and doping dependence of
the spinon mean free path
\cite{Chernyshev2005,Louis2006,Orignac2003,Rozhkov2005,Saito2003}.
But the understanding is far from complete even in this respect.
This calls for more systematic experimental studies on clean and
disordered $S=\tfrac{1}{2}$ systems to probe the effects of
defects/impurities on the spinon transport in such systems.
Experimental work on $S=\tfrac{1}{2}$ Heisenberg chain compounds
becomes especially crucial as they offer the unique possibility to
analyze directly the effect of spinon-phonon scattering processes
on heat transport in a quantum spin system since spinon-spinon
scattering is expected to be ineffective in relaxing the heat
current in these systems.

The double spin chain compound \112 and the single spin chain
compound \213, are two of the best realizations of a
$S=\tfrac{1}{2}$ Heisenberg spin chain model, differing mainly in
the structure of the spin chains
\cite{Motoyama1996,Teske1970,Teske1969}. The materials' heat
transport parallel to the chains is known to consist of a large
contribution from spinons in addition to conventional phononic
transport \cite{Hlubek2010,Hlubek2012,Kawamata2008,
Sologubenko2000,Sologubenko2001}. Earlier, \112 has been studied
with regard to the effect of substituting Ca for Sr in small
amounts \cite{Hammerath2011,Hlubek2011}. These studies resulted in
the following observations. Firstly, an off-chain impurity like Ca
causing bond disorder in the spin chains has strong effects on the
magnetic heat transport in these chains and opens a gap in the
spin excitation spectrum at low energies. Secondly, the effect of
Ca, viz. a strong suppression of $\kappa_{\mathrm{mag}}$,
decreases upon increasing its concentration and seems to saturate
at around 10\% Ca. This saturation was proposed to be related with
the disorder-induced long distance decay of the spin-spin
correlation \cite{Hlubek2011}. In the present work, we chose the
single chain compound, \213, for our study mainly because it
contains a single spin chain as opposed to the two weakly
interacting spin chains realized in \112. Thereby we are now
probing the effect of bond disorder on spinon transport in a more
simplified chain structure that is different from the double chain
CuO$_{2}$ network in \112. We observe a gradual but significant
suppression of the spinon heat conductivity
($\kappa_{\mathrm{mag}}$) with increasing amount of Ca content
which extends up to the highest doping level of 50 \%, and thus is
fundamentally different from that observed for the double chain
compound \112. Thus, Ca, an off-chain impurity, results in a more
subtle effect than in-chain impurities \cite{Kawamata2008} thereby
making it possible to probe a large doping regime up to very high
doping levels.

From our analysis we find that Ca can be described as an effective
defect situated within the chain limiting the mean free path which
ranges from very large ($>$ 1300 lattice spacings in pure \213) to
very short ($\sim 12$ lattice spacings) at the highest doping
level. Interestingly, we also observe that at maximum
concentrations of Ca (50 \%), the heat conductivity parallel to
the chains increases almost linearly at high temperatures above
$\sim$ 100 K thereby reflecting the behavior of the intrinsic
spinon heat conductivity of a $S=\tfrac{1}{2}$ Heisenberg chain.

\section{Experimental}

Single crystals (Sr$_{1-x}$Ca$_x$)$_2$CuO$_3$ with $x$ = 0, 0.01,
0.05, 0.1 and 0.5  were grown using the travelling solvent
floating zone method. The starting materials used for preparing
polycrystalline \213 were SrCO$_3$, CuO and CaCO$_3$, each of
99.99\% purity. Measurements of the thermal conductivity in the
range of 7-300 K were performed with a standard four probe
technique wherein the temperature gradient was determined using a
differential Au/Fe-Chromel thermocouple \cite{Hess2003a}. For the
transport measurements rectangular samples with typical dimensions
of (3 $\times$ 0.5 $\times$ 0.5) mm$^3$ were cut from the crystals
for each doping level with an abrasive slurry wire saw. The
longest dimension was taken parallel to the measurement axis.

\section{Results and Discussion}

\begin{figure}
\begin{center}
\includegraphics[scale=0.5]{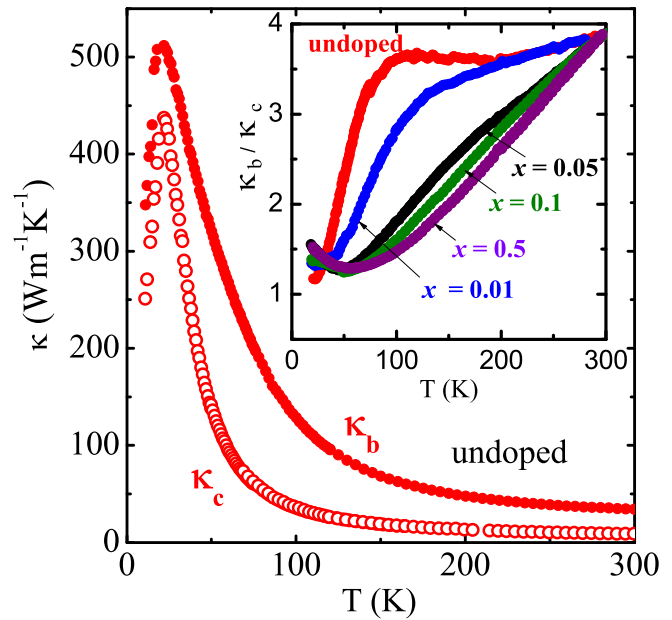}
\end{center}
\caption{(Color online) Heat conductivities $\kappa_b$ measured
parallel to the chains and $\kappa_c$ measured perpendicular to
the chains for the undoped \213 compound, as a function of
temperature (data from \cite{Hlubek2012}). Inset: Ratio of heat
conductivities $\kappa_b$ and $\kappa_c$ for \213 and the doped
compounds  to give an idea of the magnitude and temperature
dependence of the anisotropy in heat conduction. The curves for
the doped compounds are normalized to the value of
$\kappa_b$/$\kappa_c$ for the pure compound at 300 K to illustrate
the doping dependence of this ratio more clearly. The actual value
of $\kappa_b$/$\kappa_c$ at room temperature is always large
($\sim$ 4) and ranges from $\sim$ 3.8 for the pure compound to
$\sim$ 4.7 for $x$ = 0.5.} \label{Fig1}
\end{figure}

Fig.~\ref{Fig1} recalls the heat conductivity measured parallel
($\kappa_b$: solid circles) and perpendicular ($\kappa_c$: open
circles) to the spin chains of the undoped compound, as a function
of temperature $T$. A large anisotropy in the measured heat
conductivity is immediately obvious, with $\kappa_b$ along the
chains being much larger than $\kappa_c$ perpendicular to the
chains over the entire temperature range. $\kappa_c$ ($T$) shows a
sharp peak at low temperatures and falls off rather quickly at
higher temperatures, a $T$-dependence characteristic of a purely
phononic system. This heat conductivity perpendicular to the spin
chains must be purely phononic because spinons possess dispersion
only parallel to the chains. $\kappa_b$ ($T$) shows a much broader
peak at low $T$ and falls off much slower than $\kappa_c$ ($T$) at
higher temperatures, untypical of phonon-only systems. The ratio
of heat conductivities measured parallel ($\kappa_{b}$) and
perpendicular ($\kappa_{c}$) to the spin chains of undoped \213 is
plotted as a function of temperature in the inset of
Fig.~\ref{Fig1}), to give an idea of the magnitude of anisotropy
and how it varies as a function of temperature and doping. Large
anisotropy in the magnitude and temperature dependence of
$\kappa_{\parallel}$ and $\kappa_{\perp}$, anisotropic dependence
of heat conductivity on purity of compound, and unconventional
temperature dependence of $\kappa_{\parallel}$ have led to the
well established conclusion that there is an additional
contribution to the heat conduction parallel to the spin chains
due to propagating magnetic excitations called spinons
\cite{Hlubek2010,Hlubek2012,Kawamata2008,Sologubenko2000,Sologubenko2001}.
It is well confirmed that the difference between the two heat
conductivities $\kappa_{\parallel}$ and $\kappa_{\perp}$ gives a
good estimation of the spinon heat conduction in the system. This
novel channel of heat conduction in these compounds exists in
addition to the conventional channel of heat transport via
phonons.

\begin{figure}[t]
\begin{center}
\includegraphics[scale=0.5]{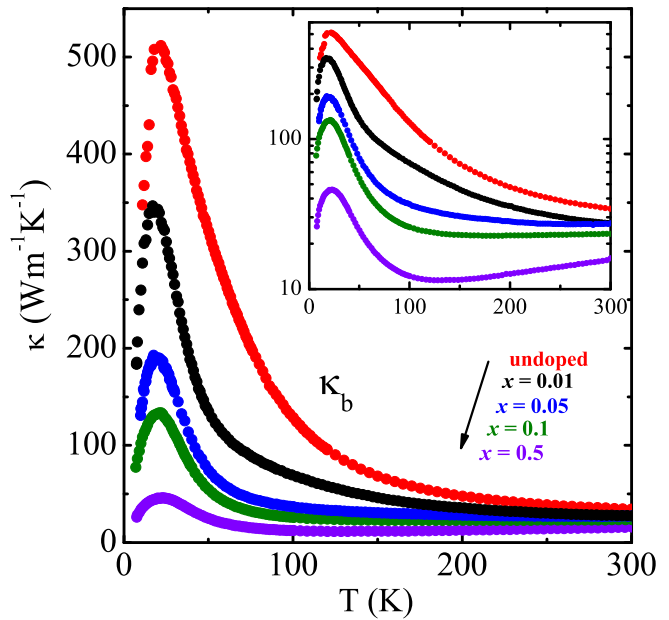}
\end{center} \caption{(Color online) Heat conductivity $\kappa_b$
as a function of temperature, measured parallel to the chains.
Inset: A semi-log plot of $\kappa_b$ versus temperature to better
illustrate the changes upon introducing disorder.} \label{Fig2}
\end{figure}

Fig.~\ref{Fig2} shows the heat conductivity of the pure and doped
compounds as a function of temperature, measured along the
direction of the spin chains ($\kappa_b$($T$)). Upon doping, the
broad peak in $\kappa_b$($T$) at $\sim$ 20 K is reduced from
$\sim$ 512 W m$^{-1}$K$^{-1}$ at $x$ = 0 to $\sim$ 350 W
m$^{-1}$K$^{-1}$ at $x$ = 0.01, and the shape of the curve
changes. The broad peak in the undoped compound develops into a
sharper phononic-like peak and a shoulder indicative of spinon
contribution that has now reduced in magnitude. Increasing the Ca
content to $x$ = 0.05 further suppresses the peak of $\kappa_b$ to
$\sim$ 195 W m$^{-1}$K$^{-1}$, and this trend continues up to $x$
= 0.5. In the inset of Fig.~\ref{Fig2}, we see that for higher
doping concentrations ($x$ = 0.1 and $x$ = 0.5) the heat
conductivity shows a positive slope above $\sim$ 100 K up to 300
K. Whereas, for lower concentrations ($x$ = 0.01 and $x$ = 0.05) a
negative slope exists up to 300 K, as in the case of the undoped
compound.

\begin{figure}[t]
\begin{center}
\includegraphics[scale=0.5]{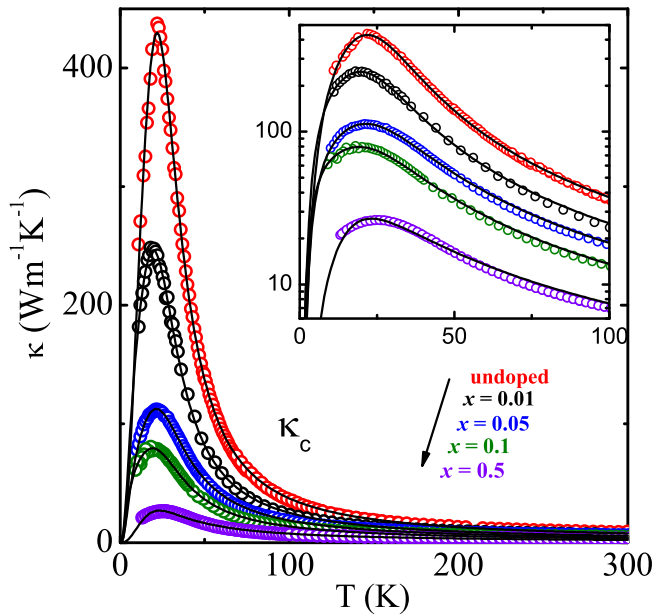}
\end{center} \caption{(Color online) Heat conductivity $\kappa$$_c$
as a function of temperature, measured perpendicular to the
chains. Inset: A semi-log plot of $\kappa_c$ versus temperature to
better illustrate the changes upon introducing disorder. The solid
black curves are fits to the curves according to the Callaway
model.} \label{Fig3}
\end{figure}

Fig.~\ref{Fig3} shows the heat conductivity measured perpendicular
to the spin chains ($\kappa_c$($T$)). Upon doping increasing
amounts of Ca, the peak magnitude of $\kappa_c$($T$) gradually
decreases as is expected for a purely phononic system due to
increased scattering of phonons off Ca impurities. The black solid
curves in Fig.~\ref{Fig3} show fits for all Ca concentrations
employing the Callaway model for phononic heat tansport
\cite{Callaway1959}. From Fig.~\ref{Fig3} we see that the fits are
good and the temperature dependence of the heat conductivity
perpendicular to the spin chains is perfectly consistent with pure
phononic heat transport. Decreasing magnitudes of $\kappa_c$ with
increasing concentrations of dopant can be captured in the model
mainly by changing the parameter related to the phonon-defect
scattering probability, thereby indicating enhanced scattering.
Details regarding these fits are given in the appendix.

\begin{figure}
\begin{center}
\includegraphics[scale=0.5]{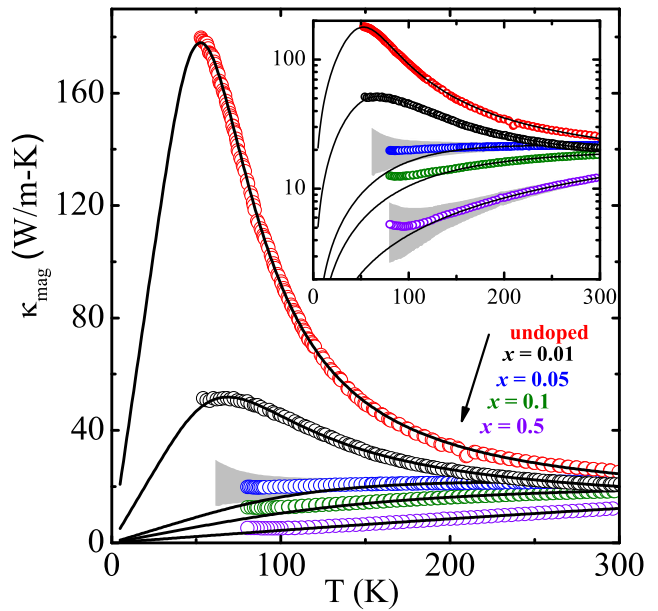}
\end{center}
\caption{(Color online) Magnetic heat conductivity
$\kappa_{\mathrm{mag}}$ as a function of temperatures for the
undoped and the Ca-doped compounds. Inset: A semi-log plot of
$\kappa_{\mathrm{mag}}$ versus temperature to more clearly show
the increase of $\kappa_{\mathrm{mag}}$ with increasing
temperature for higher concentrations of Ca. Solid black lines are
$\kappa_{\mathrm{mag}}$ recalculated using the fits for
$l_{\mathrm{mag}}$ and Eq.~\ref{eq:lmag}. The grey regions depict
the uncertainities associated with the extraction of
$\kappa$$_{mag}$.} \label{Fig4}
\end{figure}

As can be inferred from the inset of Fig.~\ref{Fig1} there is
still a considerable anisotropy present in the heat conductivity
of the Ca doped compounds signalling a still present and
significant spinon contribution. It is well known that subtracting
$\kappa_c$ from $\kappa_b$ provides a good estimate of the spinon
heat conductivity ($\kappa_{mag}$($T$))
\cite{Hlubek2010,Hlubek2012,Kawamata2008,
Sologubenko2000,Sologubenko2001}. Fig.~\ref{Fig4} shows
$\kappa_{\mathrm{mag}}$($T$) for different doping levels of Ca.
Here, we can see that $\kappa_{\mathrm{mag}}$($T$) drastically
decreases for the Ca doped samples. Strong suppression of
$\kappa_{\mathrm{mag}}$ indicates that in the doped compounds
spinon transport is gradually impeded as a result of increasing
scattering of spinons off defects induced by increasing amounts of
Ca. For higher doping concentrations, $x$ = 0.1 and 0.5, we can
clearly see that $\kappa_{\mathrm{mag}}$ increases at higher
temperatures, with an almost linear increase for $x$ = 0.1. This
interesting observation reflects the intrinsic properties of
spinon transport in a $S=\tfrac{1}{2}$ Heisenberg spin chain as
will be explained in more detail further below. The grey regions
around the curves of $x$ = 0.05 and 0.5 doped Ca depict the
uncertainty in $\kappa_{\mathrm{mag}} $\footnote{As the method for
extraction of magnetic heat conductivity relies on the assumption
of isotropic phononic heat conduction, there could always be an
error in our estimation that stems from the anisotropy of phononic
conduction parallel and perpendicular to the chains. This error is
large at low temperatures as the magnitudes of $\kappa_b$ and
$\kappa$$_c$ are large in this regime, thus creating significant
uncertainty in the extracted $\kappa_{\mathrm{mag}}$. The errors
have been calculated by taking into account a 30\% anisotropy in
the phononic heat conductivity parallel and perpendicular to the
chains.}.

We now analyze $\kappa_{\mathrm{mag}}$ by extracting the mean free
path of spinons, $l_{\mathrm{mag}}$. It can be approached in two
ways. One is by treating the low temperature spinon transport in
the framework of a semi-classical kinetic model
\cite{Hlubek2010,Hlubek2012,Kawamata2008,
Sologubenko2000,Sologubenko2001,Hess2007}, and the other is to use
the more microscopic Drude weight approach
\cite{Heidrich-Meisner2003,Heidrich-Meisner2002,
Klumper2002,Sakai2003}. In theory, intergrability of the spin-1/2
Heisenberg model results in a divergent $\kappa_{\mathrm{mag}}$
and is described by the product of the thermal Drude weight and a
delta function at zero frequency. The low temperature behaviour of
the Drude weight of a Heisenberg chain is linear in temperature
and is given as \cite{Heidrich-Meisner2003,Heidrich-Meisner2002,
Klumper2002,Sakai2003},
\begin{equation} \label{eq:drude}
D_{\mathrm{th}}=\frac{(\pi k_{B})^{2}} {3\hbar}vT,\end{equation}
where $v$ is the spinon velocity and $k_{B}$ is the Boltzmann
constant. Extrinsic scattering processes in a real system are
expected to render the thermal conductivity of a single chain
finite with a width $\sim$ $1/\tau$.\cite{Hess2007a} Thus, the
magnetic heat conductivity for a single chain can be written as
$\tilde{\kappa}_{\mathrm{mag}} = D_{\mathrm{th}} \tau / \pi$.
Combining this with the expression for the Drude weight
(Eq.~\ref{eq:drude}) and using $l_{\mathrm{mag}} = \tau v$ we get
a relation between the mean free path and the thermal conductivity
of a real system,

\begin{equation} \label{eq:lmag}
l_{\mathrm{mag}}=\frac{3\hbar}{\pi
N_{s}k_{B}^{2}T}\kappa_{\mathrm{mag}},\end{equation} where
$N_{s}=2/ab$ is the number of spin chains per unit area.

We mention that exactly the same expression can also be derived by
starting with a simple kinetic model where $\kappa_{\mathrm{mag}}$
= $\int C_{k} l_{k} v_{k}dk$. Here $C_{k}$ is the magnetic
specific heat, and $k$ denotes the momentum dependence
\cite{Hlubek2010,Hlubek2012,Kawamata2008,
Sologubenko2000,Sologubenko2001,Hess2007}. We now use
Eq.~\ref{eq:lmag} to extract and analyze the spinon mean free
path.

\begin{figure}[t]
\begin{center}
\includegraphics*[scale=0.5]{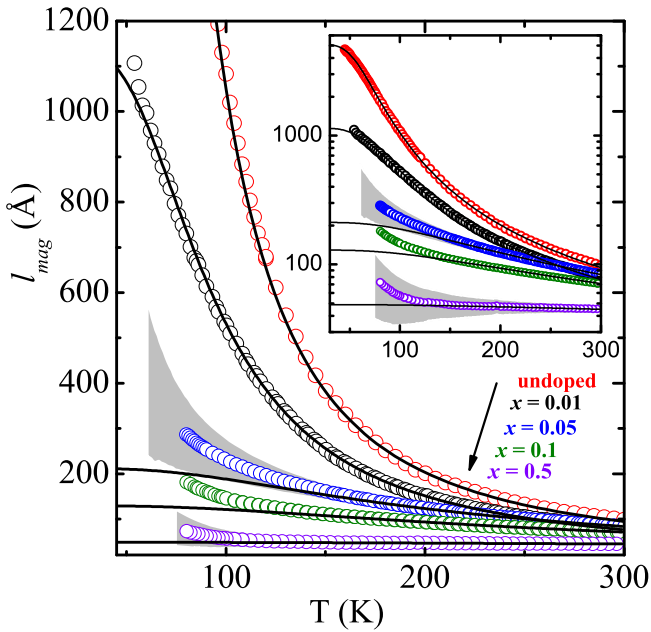}
\end{center} \caption{(Color online) Mean free path of spinons as a function of
temperature $l_{\mathrm{mag}}$($T$) and the fits according to
Eq.~\ref{eq:Umklapp}. The grey region around the curve for 1 \% Ca
depicts the uncertainity associated with the calculation of
$l_{\mathrm{mag}}$. Inset: a semi-log plot of $l_{\mathrm{mag}}$
versus temperature.} \label{Fig5}
\end{figure}

$l_{\mathrm{mag}}$ as a function of temperature for the pure and
doped compounds is plotted in Fig.~\ref{Fig5}, the inset showing a
semi-log plot of the same. Upon doping, the magnitude of the mean
free path at low temperatures strongly and monotonically
decreases, the trend being similar to that observed in
$\kappa_{\mathrm{mag}}$. In all cases, $l_\mathrm{mag}$ decreases
with increasing $T$. Note that $l_{\mathrm{mag}}$ for $x$ = 0.5 is
nearly constant at temperatures above 100 K, whereas for lower
concentrations we see a temperature dependent decrease. As has
been pointed out already for the pure compound
\cite{Hlubek2012,Kawamata2008,Sologubenko2001}, the $T$-dependent
decrease can qualitatively be well explained by spinon-phonon
scattering which becomes increasingly important with rising $T$.
The doping-induced shortening of the mean free path can be
qualitatively ascribed to strong spinon scattering due to
Ca-induced disorder in the doped compounds. We model
$l_{\mathrm{mag}}$($T$) using Matthiesen's rule and taking into
account two scattering processes for spinons, viz. spinon-defect
scattering and spinon-phonon scattering
\cite{Hlubek2010,Hlubek2012,Kawamata2008,
Sologubenko2000,Sologubenko2001,Hess2007}. Thus we have
$l_{\mathrm{mag}}^{-1}=l_{0}^{-1}+l_{\mathrm{sp}}^{-1}$, where
$l_{0}$ describes the $T$-independent spinon-defect scattering
whereas $l_{\mathrm{sp}}(T)$ accounts for the $T$-dependent
spinon-phonon scattering. For the latter, we assume a general
umklapp process with a characteristic energy scale
$k_{B}T_{u}^{*}$ of the order of the relevant phonon energies,
which is commonly used in literature
\cite{Hlubek2010,Kawamata2008,
Sologubenko2000,Sologubenko2001,Hess2007}. Thus we get the
expression,
\begin{equation}
l_{\mathrm{mag}}^{-1}=l_{0}^{-1}+\left(\frac{\exp\left(T_{u}^{*}/T\right)}{A_{s}T}\right)^{-1},\label{eq:Umklapp}\end{equation}
which can be used to fit the data with $l_{0}$, $A_{s}$ and
$T_{u}^{*}$ ($A_{s}$ is a measure of the coupling strength) as
free parameters.

\begin{figure}[t]
\begin{center}
\includegraphics[scale=0.5]{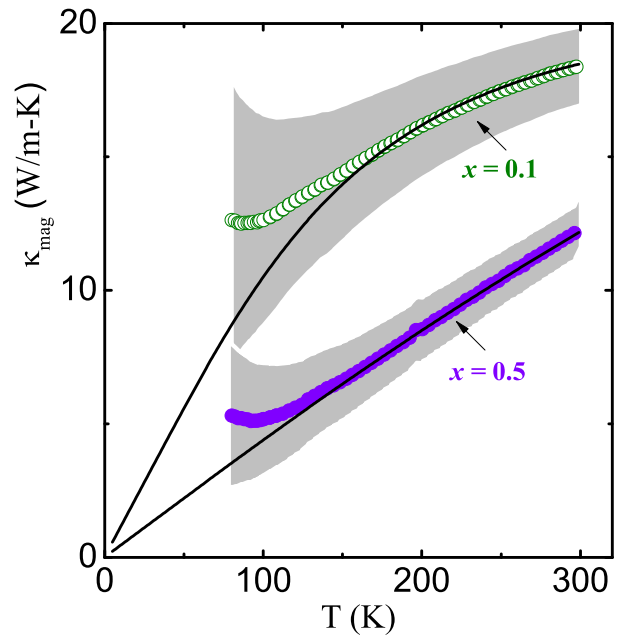}
\end{center} \caption{(Color online) Magnetic heat conductivity
$\kappa_{\mathrm{mag}}$ as a function of temperatures for the 10
\% and 50 \% Ca doped compounds. Solid lines are
$\kappa_{\mathrm{mag}}$ recalculated using the fits for
$l_{\mathrm{mag}}$ and Eq.~\ref{eq:lmag}. Grey regions around the
curve indicate the uncertainties in $\kappa_{\mathrm{mag}}$}
 \label{Fig6}
\end{figure}

The fits for the undoped and doped compounds are shown in
Fig.~\ref{Fig5} and the fit parameters in Table~\ref{tab1}. We
obtain good fits for the pure and $x$ = 0.01 compounds in the
temperature range from 50-300 K, and for the $x$ = 0.05, 0.1, 0.5
compounds the fits are good in the temperature range from 150-300
K. It is apparently enough to take into account just spinon-defect
scattering to describe the doping dependence of $l_{\mathrm{mag}}$
and umklapp-like spinon-phonon scattering to explain the
$T$-dependence of $l_{\mathrm{mag}}$, where the latter remains
essentially the same upon doping. In order to fit the data, we
have fixed $T_{u}^{*}$ as 210 K (obtained by fitting
$l_{\mathrm{mag}}$ for the pure compound) for the whole doping
series allowing $l_{0}$ and $A_{s}$ to vary. In the model,
$T_{u}^{*}$ stems from the characteristic energy scale
$k_{B}T_{u}^{*}$ of the umklapp process, which is of the order of
the Debye energy \cite{Sologubenko2000,Sologubenko2001} suggestive
of acoustic phonons being involved in the scattering process.
Hence, it is physically justified to fix $T_{u}^{*}$ for the doped
compounds, as the Debye energy is not expected to change
substantially with doping. $A_{s}$ changes slightly as a function
of doping but is of the same order of magnitude for all doping
levels. The slight variation in this parameter accounts for
uncertainties in determining $\kappa_{\mathrm{mag}}$, and a small
variation in the spin-coupling strength in the pure and doped
compounds.

The spinon-defect scattering length $l_{0}$, which represents a
lower bound for the low-$T$ limit of $l_{\mathrm{mag}}$ and which
should significantly depend on the sample's purity, decreases
strongly upon doping and is very different for the pure and the
doped compounds as can be seen in Table~\ref{tab1}. This parameter
is most sensitive to changes in the Ca concentration, indicating
that Ca defects primarily act as efficient barriers for the
propagating spinons, i.e the disorder-induced scattering can be
well described by intra-chain defects. We mention that the
deviation of the experimentally extracted $l_{\mathrm{mag}}$ from
the fits at low temperatures ($\lesssim$ 150 K) can be attributed
to the large error inherent in $l_{\mathrm{mag}}$ which is shown
by the grey region surrounding the curves. As this error is large
for the heavily doped compounds where the mean free paths are
small, the fits are expected to be inaccurate at low temperatures.

\begin{table}[hb]
    \centering
    \caption{Fit parameters obtained by fitting $\l_{mag}$ by Eq.~\ref{eq:Umklapp}.}
    \label{tab:fitparameters}
\begin{tabular}{ccccc}
\hline \hline
 Ca          & $\l_0$                   & $T^*_u$      & $A_s$  \\
 content     & (\AA)                   & (K)          & ($10^{5}$ m/K) \\
 \hline
 0 \%        & 5093                   & 210        & 7.39     \\

 1 \%        & 1139                     & 210        & 8.19   \\

 5 \%        & 212                     & 210        & 4.85  \\

10 \%        & 127                     & 210        & 4.2  \\

50 \%        & 48                     & 210        & 1.19  \\
 \hline \hline
\end{tabular}
 \label{tab1}

\end{table}

Using the fits obtained for $l_{\mathrm{mag}}$, we can recalculate
$\kappa_{\mathrm{mag}}$ using Eq.~\ref{eq:lmag}, and plot these
curves (black solid curves in Fig.~\ref{Fig4}) over the
experimental curves to have a further illustration of the
analysis. These recalculated curves, within the kinetic model,
give us a good idea of the evolution of $\kappa_{\mathrm{mag}}$ in
the entire temperature range, from the dilute doped compound,
where it smoothly decreases with increasing temperatures, to the
heavily doped compound where it increases almost linearly with
increasing temperatures.

We now turn in more detail to $\kappa_{\mathrm{mag}}$ for the $x$
= 0.1 and $x$ = 0.5 doped compounds for which a monotonic increase
with rising $T$ is found (Fig.~\ref{Fig6}). Remarkably, an almost
linear temperature dependence above $T$ $\approx$ 150 K is seen
for $x$ = 0.5. Such a linear temperature dependence of
$\kappa_{\mathrm{mag}}$ has been observed before in the highly
disordered quasi one-dimensional compound CaCu$_{2}$O$_{3}$ and
reveals the intrinsic temperature dependence of the heat transport
of a $S=\tfrac{1}{2}$ Heisenberg chain \cite{Hess2007a}. It is
well known that the temperature independent spinon-defect
scattering mechanism that leads to a temperature independent
$l_{\mathrm{mag}}$ is the dominant mechanism at low temperatures
and at higher temperatures spinon-phonon scattering becomes
dominant. Having large number of defects in the chain, which is
the case for the doped compounds, will enhance the probability of
spinons scattering off defects over that of spinons scattering off
phonons. If the defect concentration is sufficiently high, the
temperature dependence of the spinon-phonon scattering mechanism
can be completely masked by the temperature independent
spinon-defect scattering mechanism and, in turn, lead to a
temperature independent spinon mean free path, i.e. a temperature
independent scattering time $\tau$. Thus, the $T$-dependence of
the experimental $\kappa_{\mathrm{mag}}$ represents directly that
of the thermal Drude weight (Eq.~\ref{eq:drude}).

\begin{figure}
\begin{center}
\includegraphics[scale=0.45]{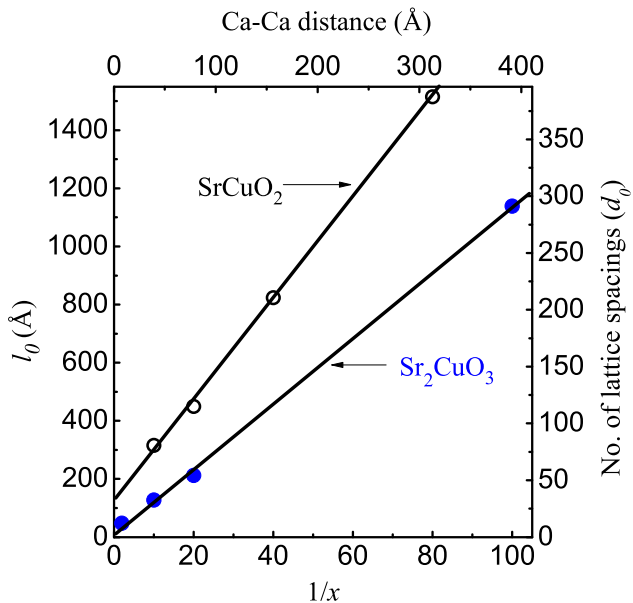}
\end{center} \caption{(Color online) Spinon-defect scattering length, $l_{0}$, is
plotted against the mean distance between two Ca atoms (lower
x-axis) and the inverse of Ca concentration ($x$) (upper x-axis)
for \112 (open symbols) and \213 (filled symbols); the solid lines
are linear fits to the data. The y-axis on the right gives an idea
of the mean free path in terms of the number of lattice spacings
between two Cu sites($d_0$).} \label{Fig7}
\end{figure}

Finally, in order to investigate the Ca-induced scattering process
further, we plot the obtained values of the spinon-defect
scattering length $l_{0}$ as a function of the mean distance
between two Ca atoms and the inverse of Ca concentration in
Fig.~\ref{Fig7}. Here, we see that $l_{0}$ scales perfectly with
the inverse of Ca concentration as $l_{0}$ = 2.93(1/$x$)\AA
$\sim$3$d_0$, where $d_0=3.91$ is the lattice spacing between two
Cu sites along the chain \cite{Teske1970,Teske1969}. This
corroborates that the Ca-induced bond disorder can perfectly be
captured in terms of effective defects in the chain, where the
scattering probability per defect is equally strong at all
concentrations. From this plot we also see that the Ca-Ca distance
is smaller by a factor of 3 than the spinon-defect scattering
length $l_{0}$.

We compare this finding with the results of a recent analogous
study of the Ca-doped double spin chain compound, \112
\cite{Hlubek2011} (open symbols in Fig.~\ref{Fig7}). For this
compound we see that $l_{0}$ can be described as $l_{0} =
4.3(1/x)$\AA + $l_{lim}$, where $l_{lim}$ ($\approx$ 140\AA) is an
offset. Therefore, the spinon-defect scattering length, $l_{0}$,
does not scale with the inverse of the Ca concentration,
indicating that already at intermediate concentrations ($x$ = 0.1)
the effect of Ca saturates and the mean free path of spinons is
not reduced any further upon increasing the Ca concentration. This
is different from the perfect scaling that we observe for the
single chain compound. The offset observed in Ca-doped \112 was
interpreted to be due to a limit set by the disorder-induced long
distance decay of the spin-spin correlation ($\xi$), $\xi$ being
calculated for a single $S=\tfrac{1}{2}$ Heisenberg chain
\cite{Hlubek2011}. However, this claim does not seem to be valid
in our case as the effect of Ca is equally strong at all doping
levels. Thus, the proposed connection between the two quantities
ought to be just a coincidence. We thus conclude that in the two
compounds the difference in the effect of Ca-induced bond disorder
could be in some way related to the difference in the chain
structure. \112 has a double chain structure with a finite
interchain coupling as opposed to the single chain structure in
\213. Although this coupling is small compared to the coupling
strength within each chain, it seems to play an important role in
deciding the effect of bond-disorder on the spinons propagating in
these chains. This is evident as there is no other significant
difference between these two compounds with regard to the spin
system.

\section{Conclusion}

We have studied the effect of introducing an off-chain impurity
like Ca, thereby creating bond disorder, on the heat transport of
the prototype single spin chain compound \213. We observe a
drastic suppression of the magnetic heat conductivity parallel to
the chains indicating that the propagation of spinons is very
sensitive to even the slightest bond disorder. The temperature
dependence of the mean free path of spinons can be modelled by
spinon-defect and spinon-phonon scattering processes, and the
reduction of the mean free path upon doping is accounted for
mainly by increased scattering of spinons off effective in-chain
defects, where the scattering probability per defect is equally
strong in the entire doping range. This result is very different
from the case of Ca-doped double spin chain compound \112 where,
presumably, finite inter-chain interaction reduces the effect of
Ca at higher doping levels.

Interestingly, large disorder present in the compounds doped with
high concentrations of Ca leads to a linearly increasing intrinsic
spinon heat transport of the spin chain due to prevailing
spinon-defect scattering and thus a vanishing temperature
dependence of the spinon mean free path. Thus, Ca-doped \213
represents a unique case where the impact of impurities can be
studied in a wide range of doping, covering very clean as well as
very dirty limits.

\section*{Appendix}

\begin{table}[hb]
    \centering
    \caption{Callaway fit parameters for the $\kappa$$_c$($T$) curves.}
    \label{tab:fitparameters}
\begin{tabular}{cccccc}
\hline \hline
 Ca          &     B                & b              &      A             &      L          &   $\Theta$$_D$  \\
 content     & ($10^{-31}$ K$^{-1}$ s$^2$)         &                & ($10^{-44}$ s$^3$)       &  ($10^{-4}$ m)    &     (K)         \\
 \hline
 0 \%        & 3.55                 & 2.41           & 0.31                &   2.98         & 261 K           \\

 1 \%        & 5.07                 & 2.80           & 1.42               &   7.82          & 265 K           \\

 5 \%        & 6.13                 & 2.84           & 3.69               &   4.72          & 262 K           \\

 10 \%       & 8.35                 & 3.13           & 9.77               &   10.7          & 262 K           \\

 50 \%       & 8.60                 & 4.36           & 13.82               &   0.17          & 265 K           \\
 \hline \hline
\end{tabular}
\label{tab2}

\end{table}

We model the phononic heat conductivity perpendicular to the
chains using a phenomenological model devised by Callaway
\cite{Callaway1959} using,
\begin{equation}
\kappa_{\mathrm{ph}}=\frac{k_{B}}{2\pi^{2}v_{\mathrm{ph}}}\left(\frac{k_{B}T}{\hbar}\right)^{3}\int_{0}^{\Theta_{D}/T}\frac{x^{4}\mathrm{e}^{x}}{\left(\mathrm{e}^{x}-1\right)^{2}}\cdot\tau_{c}\mathrm{d}x.\end{equation}
Here $x=\hbar\omega/k_{B}T$, $\omega$ is the phonon angular
frequency, $\Theta_{D}$ is the Debye temperature and
$v_{\mathrm{ph}}$ is the phonon velocity. $\tau_{c}$ is a combined
scattering rate, which is assumed to be the sum of individual
scattering rates,
\begin{equation}
\tau_{c}^{-1}=\tau_{B}^{-1}+\tau_{D}^{-1}+\tau_{U}^{-1},\end{equation}
where $\tau_{B}$ denotes boundary scattering, $\tau_{D}$ point
defect scattering and $\tau_{U}$ Umklapp scattering. Including the
expressions for each scattering process, $\tau_{c}$ can then be
written as \begin{equation}
\tau_{c}^{-1}=\frac{v_{\mathrm{ph}}}{L}+A\omega^{4}+B\omega^{2}T\exp\left(-\frac{\Theta_{D}}{bT}\right),\end{equation}
with fit parameters $L$, $A$, $B$, $b$. $A$ describes the
concentration of point defects, $L$ describes the boundary
scattering, $B$ and $b$ are intra-phonon scattering parameters.
Fig.~\ref{Fig3} shows $\kappa_c$ ($T$) for all Ca concentrations
and the corresponding Callaway fits. The obtained values of the
free parameters from the best fit are given in Table~\ref{tab2}.
$\kappa_c$ ($T$) for the pure compound was fit first, and then
$\kappa_c$ ($T$) for the doped compounds were tried to be fit by
keeping all other parameters except A fixed, i.e. only varying the
phonon-defect scattering strength. It was found that it is not
possible to fit the $\kappa_c$ ($T$) curves just by changing this
parameter. To obtain good fits, the intra-phonon scattering
parameters and the boundary scattering length also had to be
varied. One can see that the parameter $A$ steadily increases with
increasing concentration of Ca defects indicating the increasing
strength of phonon-defect scattering, which is mainly responsible
for the reduction of $\kappa_c$ ($T$) upon doping. The difference
in boundary scattering parameters only indicates a difference in
the sample geometries. Finally, a small variation in parameter $B$
could imply that the dopant possibly induces slight changes in the
phononic dispersion branches, thereby affecting the scattering
strength between phonons, in addition to playing the role of
defects-like scatterers.

\begin{acknowledgments}
This work has been supported by the European Commission through
the LOTHERM project (Project No. PITN-GA-2009-238475, and by
Deutsche Forschungsgemeinschaft (DFG) through FOR912 (HE3439/8)
and through the D-A-CH project HE 3439/12.
\end{acknowledgments}

\bibliography{Cadoped213}

\begin{thebibliography}{35}%
\makeatletter
\providecommand \@ifxundefined [1]{%
 \@ifx{#1\undefined}
}%
\providecommand \@ifnum [1]{%
 \ifnum #1\expandafter \@firstoftwo
 \else \expandafter \@secondoftwo
 \fi
}%
\providecommand \@ifx [1]{%
 \ifx #1\expandafter \@firstoftwo
 \else \expandafter \@secondoftwo
 \fi
}%
\providecommand \natexlab [1]{#1}%
\providecommand \enquote  [1]{``#1''}%
\providecommand \bibnamefont  [1]{#1}%
\providecommand \bibfnamefont [1]{#1}%
\providecommand \citenamefont [1]{#1}%
\providecommand \href@noop [0]{\@secondoftwo}%
\providecommand \href [0]{\begingroup \@sanitize@url \@href}%
\providecommand \@href[1]{\@@startlink{#1}\@@href}%
\providecommand \@@href[1]{\endgroup#1\@@endlink}%
\providecommand \@sanitize@url [0]{\catcode `\\12\catcode `\$12\catcode
  `\&12\catcode `\#12\catcode `\^12\catcode `\_12\catcode `\%12\relax}%
\providecommand \@@startlink[1]{}%
\providecommand \@@endlink[0]{}%
\providecommand \url  [0]{\begingroup\@sanitize@url \@url }%
\providecommand \@url [1]{\endgroup\@href {#1}{\urlprefix }}%
\providecommand \urlprefix  [0]{URL }%
\providecommand \Eprint [0]{\href }%
\providecommand \doibase [0]{http://dx.doi.org/}%
\providecommand \selectlanguage [0]{\@gobble}%
\providecommand \bibinfo  [0]{\@secondoftwo}%
\providecommand \bibfield  [0]{\@secondoftwo}%
\providecommand \translation [1]{[#1]}%
\providecommand \BibitemOpen [0]{}%
\providecommand \bibitemStop [0]{}%
\providecommand \bibitemNoStop [0]{.\EOS\space}%
\providecommand \EOS [0]{\spacefactor3000\relax}%
\providecommand \BibitemShut  [1]{\csname bibitem#1\endcsname}%
\let\auto@bib@innerbib\@empty
\bibitem [{\citenamefont {Sologubenko}\ \emph {et~al.}(2000)\citenamefont
  {Sologubenko}, \citenamefont {Felder}, \citenamefont {Giann\`o},
  \citenamefont {Ott}, \citenamefont {Vietkine},\ and\ \citenamefont
  {Revcolevschi}}]{Sologubenko2000}%
  \BibitemOpen
  \bibfield  {author} {\bibinfo {author} {\bibfnamefont {A.~V.}\ \bibnamefont
  {Sologubenko}}, \bibinfo {author} {\bibfnamefont {E.}~\bibnamefont {Felder}},
  \bibinfo {author} {\bibfnamefont {K.}~\bibnamefont {Giann\`o}}, \bibinfo
  {author} {\bibfnamefont {H.~R.}\ \bibnamefont {Ott}}, \bibinfo {author}
  {\bibfnamefont {A.}~\bibnamefont {Vietkine}}, \ and\ \bibinfo {author}
  {\bibfnamefont {A.}~\bibnamefont {Revcolevschi}},\ }\href {\doibase
  10.1103/PhysRevB.62.R6108} {\bibfield  {journal} {\bibinfo  {journal} {Phys.
  Rev. B}\ }\textbf {\bibinfo {volume} {62}},\ \bibinfo {pages} {R6108}
  (\bibinfo {year} {2000})}\BibitemShut {NoStop}%
\bibitem [{\citenamefont {Sologubenko}\ \emph {et~al.}(2001)\citenamefont
  {Sologubenko}, \citenamefont {Giann\`o}, \citenamefont {Ott}, \citenamefont
  {Vietkine},\ and\ \citenamefont {Revcolevschi}}]{Sologubenko2001}%
  \BibitemOpen
  \bibfield  {author} {\bibinfo {author} {\bibfnamefont {A.~V.}\ \bibnamefont
  {Sologubenko}}, \bibinfo {author} {\bibfnamefont {K.}~\bibnamefont
  {Giann\`o}}, \bibinfo {author} {\bibfnamefont {H.~R.}\ \bibnamefont {Ott}},
  \bibinfo {author} {\bibfnamefont {A.}~\bibnamefont {Vietkine}}, \ and\
  \bibinfo {author} {\bibfnamefont {A.}~\bibnamefont {Revcolevschi}},\ }\href
  {\doibase 10.1103/PhysRevB.64.054412} {\bibfield  {journal} {\bibinfo
  {journal} {Phys. Rev. B}\ }\textbf {\bibinfo {volume} {64}},\ \bibinfo
  {pages} {054412} (\bibinfo {year} {2001})}\BibitemShut {NoStop}%
\bibitem [{\citenamefont {Hlubek}\ \emph {et~al.}(2011)\citenamefont {Hlubek},
  \citenamefont {Ribeiro}, \citenamefont {Saint-Martin}, \citenamefont
  {Nishimoto}, \citenamefont {Revcolevschi}, \citenamefont {Drechsler},
  \citenamefont {Behr}, \citenamefont {Trinckauf}, \citenamefont
  {Hamann-Borrero}, \citenamefont {Geck}, \citenamefont {B\"uchner},\ and\
  \citenamefont {Hess}}]{Hlubek2011}%
  \BibitemOpen
  \bibfield  {author} {\bibinfo {author} {\bibfnamefont {N.}~\bibnamefont
  {Hlubek}}, \bibinfo {author} {\bibfnamefont {P.}~\bibnamefont {Ribeiro}},
  \bibinfo {author} {\bibfnamefont {R.}~\bibnamefont {Saint-Martin}}, \bibinfo
  {author} {\bibfnamefont {S.}~\bibnamefont {Nishimoto}}, \bibinfo {author}
  {\bibfnamefont {A.}~\bibnamefont {Revcolevschi}}, \bibinfo {author}
  {\bibfnamefont {S.-L.}\ \bibnamefont {Drechsler}}, \bibinfo {author}
  {\bibfnamefont {G.}~\bibnamefont {Behr}}, \bibinfo {author} {\bibfnamefont
  {J.}~\bibnamefont {Trinckauf}}, \bibinfo {author} {\bibfnamefont {J.~E.}\
  \bibnamefont {Hamann-Borrero}}, \bibinfo {author} {\bibfnamefont
  {J.}~\bibnamefont {Geck}}, \bibinfo {author} {\bibfnamefont {B.}~\bibnamefont
  {B\"uchner}}, \ and\ \bibinfo {author} {\bibfnamefont {C.}~\bibnamefont
  {Hess}},\ }\href {\doibase 10.1103/PhysRevB.84.214419} {\bibfield  {journal}
  {\bibinfo  {journal} {Phys. Rev. B}\ }\textbf {\bibinfo {volume} {84}},\
  \bibinfo {pages} {214419} (\bibinfo {year} {2011})}\BibitemShut {NoStop}%
\bibitem [{\citenamefont {Hlubek}\ \emph {et~al.}(2010)\citenamefont {Hlubek},
  \citenamefont {Ribeiro}, \citenamefont {Saint-Martin}, \citenamefont
  {Revcolevschi}, \citenamefont {Roth}, \citenamefont {Behr}, \citenamefont
  {B\"uchner},\ and\ \citenamefont {Hess}}]{Hlubek2010}%
  \BibitemOpen
  \bibfield  {author} {\bibinfo {author} {\bibfnamefont {N.}~\bibnamefont
  {Hlubek}}, \bibinfo {author} {\bibfnamefont {P.}~\bibnamefont {Ribeiro}},
  \bibinfo {author} {\bibfnamefont {R.}~\bibnamefont {Saint-Martin}}, \bibinfo
  {author} {\bibfnamefont {A.}~\bibnamefont {Revcolevschi}}, \bibinfo {author}
  {\bibfnamefont {G.}~\bibnamefont {Roth}}, \bibinfo {author} {\bibfnamefont
  {G.}~\bibnamefont {Behr}}, \bibinfo {author} {\bibfnamefont {B.}~\bibnamefont
  {B\"uchner}}, \ and\ \bibinfo {author} {\bibfnamefont {C.}~\bibnamefont
  {Hess}},\ }\href {\doibase 10.1103/PhysRevB.81.020405} {\bibfield  {journal}
  {\bibinfo  {journal} {Phys. Rev. B}\ }\textbf {\bibinfo {volume} {81}},\
  \bibinfo {pages} {020405} (\bibinfo {year} {2010})}\BibitemShut {NoStop}%
\bibitem [{\citenamefont {Hlubek}\ \emph {et~al.}(2012)\citenamefont {Hlubek},
  \citenamefont {Zotos}, \citenamefont {Singh}, \citenamefont {Saint-Martin},
  \citenamefont {Revcolevschi}, \citenamefont {BÃ¼chner},\ and\ \citenamefont
  {Hess}}]{Hlubek2012}%
  \BibitemOpen
  \bibfield  {author} {\bibinfo {author} {\bibfnamefont {N.}~\bibnamefont
  {Hlubek}}, \bibinfo {author} {\bibfnamefont {X.}~\bibnamefont {Zotos}},
  \bibinfo {author} {\bibfnamefont {S.}~\bibnamefont {Singh}}, \bibinfo
  {author} {\bibfnamefont {R.}~\bibnamefont {Saint-Martin}}, \bibinfo {author}
  {\bibfnamefont {A.}~\bibnamefont {Revcolevschi}}, \bibinfo {author}
  {\bibfnamefont {B.}~\bibnamefont {BÃ¼chner}}, \ and\ \bibinfo {author}
  {\bibfnamefont {C.}~\bibnamefont {Hess}},\ }\href
  {http://stacks.iop.org/1742-5468/2012/i=03/a=P03006} {\bibfield  {journal}
  {\bibinfo  {journal} {Journal of Statistical Mechanics: Theory and
  Experiment}\ }\textbf {\bibinfo {volume} {2012}},\ \bibinfo {pages} {P03006}
  (\bibinfo {year} {2012})}\BibitemShut {NoStop}%
\bibitem [{\citenamefont {Kawamata}\ \emph {et~al.}(2008)\citenamefont
  {Kawamata}, \citenamefont {Takahashi}, \citenamefont {Adachi}, \citenamefont
  {Noji}, \citenamefont {Kudo}, \citenamefont {Kobayashi},\ and\ \citenamefont
  {Koike}}]{Kawamata2008}%
  \BibitemOpen
  \bibfield  {author} {\bibinfo {author} {\bibfnamefont {T.}~\bibnamefont
  {Kawamata}}, \bibinfo {author} {\bibfnamefont {N.}~\bibnamefont {Takahashi}},
  \bibinfo {author} {\bibfnamefont {T.}~\bibnamefont {Adachi}}, \bibinfo
  {author} {\bibfnamefont {T.}~\bibnamefont {Noji}}, \bibinfo {author}
  {\bibfnamefont {K.}~\bibnamefont {Kudo}}, \bibinfo {author} {\bibfnamefont
  {N.}~\bibnamefont {Kobayashi}}, \ and\ \bibinfo {author} {\bibfnamefont
  {Y.}~\bibnamefont {Koike}},\ }\href {\doibase 10.1143/JPSJ.77.034607}
  {\bibfield  {journal} {\bibinfo  {journal} {Journal of the Physical Society
  of Japan}\ }\textbf {\bibinfo {volume} {77}},\ \bibinfo {pages} {034607}
  (\bibinfo {year} {2008})}\BibitemShut {NoStop}%
\bibitem [{\citenamefont {Zotos}(1999)}]{Zotos1999}%
  \BibitemOpen
  \bibfield  {author} {\bibinfo {author} {\bibfnamefont {X.}~\bibnamefont
  {Zotos}},\ }\href {\doibase 10.1103/PhysRevLett.82.1764} {\bibfield
  {journal} {\bibinfo  {journal} {Phys. Rev. Lett.}\ }\textbf {\bibinfo
  {volume} {82}},\ \bibinfo {pages} {1764} (\bibinfo {year}
  {1999})}\BibitemShut {NoStop}%
\bibitem [{\citenamefont {Alvarez}\ and\ \citenamefont
  {Gros}(2002)}]{Alvarez2002}%
  \BibitemOpen
  \bibfield  {author} {\bibinfo {author} {\bibfnamefont {J.~V.}\ \bibnamefont
  {Alvarez}}\ and\ \bibinfo {author} {\bibfnamefont {C.}~\bibnamefont {Gros}},\
  }\href {\doibase 10.1103/PhysRevLett.89.156603} {\bibfield  {journal}
  {\bibinfo  {journal} {Phys. Rev. Lett.}\ }\textbf {\bibinfo {volume} {89}},\
  \bibinfo {pages} {156603} (\bibinfo {year} {2002})}\BibitemShut {NoStop}%
\bibitem [{\citenamefont {Azuma}\ \emph {et~al.}(1998)\citenamefont {Azuma},
  \citenamefont {Takano},\ and\ \citenamefont {Eccleston}}]{Azuma1998}%
  \BibitemOpen
  \bibfield  {author} {\bibinfo {author} {\bibfnamefont {M.}~\bibnamefont
  {Azuma}}, \bibinfo {author} {\bibfnamefont {M.}~\bibnamefont {Takano}}, \
  and\ \bibinfo {author} {\bibfnamefont {R.~S.}\ \bibnamefont {Eccleston}},\
  }\href {\doibase 10.1143/JPSJ.67.740} {\bibfield  {journal} {\bibinfo
  {journal} {Journal of the Physical Society of Japan}\ }\textbf {\bibinfo
  {volume} {67}},\ \bibinfo {pages} {740} (\bibinfo {year} {1998})}\BibitemShut
  {NoStop}%
\bibitem [{\citenamefont {Castella}\ \emph {et~al.}(1995)\citenamefont
  {Castella}, \citenamefont {Zotos},\ and\ \citenamefont
  {Prelov\ifmmode~\check{s}\else \v{s}\fi{}ek}}]{Castella1995}%
  \BibitemOpen
  \bibfield  {author} {\bibinfo {author} {\bibfnamefont {H.}~\bibnamefont
  {Castella}}, \bibinfo {author} {\bibfnamefont {X.}~\bibnamefont {Zotos}}, \
  and\ \bibinfo {author} {\bibfnamefont {P.}~\bibnamefont
  {Prelov\ifmmode~\check{s}\else \v{s}\fi{}ek}},\ }\href {\doibase
  10.1103/PhysRevLett.74.972} {\bibfield  {journal} {\bibinfo  {journal} {Phys.
  Rev. Lett.}\ }\textbf {\bibinfo {volume} {74}},\ \bibinfo {pages} {972}
  (\bibinfo {year} {1995})}\BibitemShut {NoStop}%
\bibitem [{\citenamefont {Chernyshev}\ and\ \citenamefont
  {Rozhkov}(2005)}]{Chernyshev2005}%
  \BibitemOpen
  \bibfield  {author} {\bibinfo {author} {\bibfnamefont {A.~L.}\ \bibnamefont
  {Chernyshev}}\ and\ \bibinfo {author} {\bibfnamefont {A.~V.}\ \bibnamefont
  {Rozhkov}},\ }\href {\doibase 10.1103/PhysRevB.72.104423} {\bibfield
  {journal} {\bibinfo  {journal} {Phys. Rev. B}\ }\textbf {\bibinfo {volume}
  {72}},\ \bibinfo {pages} {104423} (\bibinfo {year} {2005})}\BibitemShut
  {NoStop}%
\bibitem [{\citenamefont {Heidrich-Meisner}\ \emph {et~al.}(2003)\citenamefont
  {Heidrich-Meisner}, \citenamefont {Honecker}, \citenamefont {Cabra},\ and\
  \citenamefont {Brenig}}]{Heidrich-Meisner2003}%
  \BibitemOpen
  \bibfield  {author} {\bibinfo {author} {\bibfnamefont {F.}~\bibnamefont
  {Heidrich-Meisner}}, \bibinfo {author} {\bibfnamefont {A.}~\bibnamefont
  {Honecker}}, \bibinfo {author} {\bibfnamefont {D.~C.}\ \bibnamefont {Cabra}},
  \ and\ \bibinfo {author} {\bibfnamefont {W.}~\bibnamefont {Brenig}},\ }\href
  {\doibase 10.1103/PhysRevB.68.134436} {\bibfield  {journal} {\bibinfo
  {journal} {Phys. Rev. B}\ }\textbf {\bibinfo {volume} {68}},\ \bibinfo
  {pages} {134436} (\bibinfo {year} {2003})}\BibitemShut {NoStop}%
\bibitem [{\citenamefont {Heidrich-Meisner}\ \emph {et~al.}(2002)\citenamefont
  {Heidrich-Meisner}, \citenamefont {Honecker}, \citenamefont {Cabra},\ and\
  \citenamefont {Brenig}}]{Heidrich-Meisner2002}%
  \BibitemOpen
  \bibfield  {author} {\bibinfo {author} {\bibfnamefont {F.}~\bibnamefont
  {Heidrich-Meisner}}, \bibinfo {author} {\bibfnamefont {A.}~\bibnamefont
  {Honecker}}, \bibinfo {author} {\bibfnamefont {D.~C.}\ \bibnamefont {Cabra}},
  \ and\ \bibinfo {author} {\bibfnamefont {W.}~\bibnamefont {Brenig}},\ }\href
  {\doibase 10.1103/PhysRevB.66.140406} {\bibfield  {journal} {\bibinfo
  {journal} {Phys. Rev. B}\ }\textbf {\bibinfo {volume} {66}},\ \bibinfo
  {pages} {140406} (\bibinfo {year} {2002})}\BibitemShut {NoStop}%
\bibitem [{\citenamefont {Hess}\ \emph
  {et~al.}(2003{\natexlab{a}})\citenamefont {Hess}, \citenamefont {B\"uchner},
  \citenamefont {Ammerahl}, \citenamefont {Colonescu}, \citenamefont
  {Heidrich-Meisner}, \citenamefont {Brenig},\ and\ \citenamefont
  {Revcolevschi}}]{Hess2003}%
  \BibitemOpen
  \bibfield  {author} {\bibinfo {author} {\bibfnamefont {C.}~\bibnamefont
  {Hess}}, \bibinfo {author} {\bibfnamefont {B.}~\bibnamefont {B\"uchner}},
  \bibinfo {author} {\bibfnamefont {U.}~\bibnamefont {Ammerahl}}, \bibinfo
  {author} {\bibfnamefont {L.}~\bibnamefont {Colonescu}}, \bibinfo {author}
  {\bibfnamefont {F.}~\bibnamefont {Heidrich-Meisner}}, \bibinfo {author}
  {\bibfnamefont {W.}~\bibnamefont {Brenig}}, \ and\ \bibinfo {author}
  {\bibfnamefont {A.}~\bibnamefont {Revcolevschi}},\ }\href {\doibase
  10.1103/PhysRevLett.90.197002} {\bibfield  {journal} {\bibinfo  {journal}
  {Phys. Rev. Lett.}\ }\textbf {\bibinfo {volume} {90}},\ \bibinfo {pages}
  {197002} (\bibinfo {year} {2003}{\natexlab{a}})}\BibitemShut {NoStop}%
\bibitem [{\citenamefont {Hess}\ \emph
  {et~al.}(2003{\natexlab{b}})\citenamefont {Hess}, \citenamefont {B\"uchner},
  \citenamefont {Ammerahl},\ and\ \citenamefont {Revcolevschi}}]{Hess2003a}%
  \BibitemOpen
  \bibfield  {author} {\bibinfo {author} {\bibfnamefont {C.}~\bibnamefont
  {Hess}}, \bibinfo {author} {\bibfnamefont {B.}~\bibnamefont {B\"uchner}},
  \bibinfo {author} {\bibfnamefont {U.}~\bibnamefont {Ammerahl}}, \ and\
  \bibinfo {author} {\bibfnamefont {A.}~\bibnamefont {Revcolevschi}},\ }\href
  {\doibase 10.1103/PhysRevB.68.184517} {\bibfield  {journal} {\bibinfo
  {journal} {Phys. Rev. B}\ }\textbf {\bibinfo {volume} {68}},\ \bibinfo
  {pages} {184517} (\bibinfo {year} {2003}{\natexlab{b}})}\BibitemShut
  {NoStop}%
\bibitem [{\citenamefont {Hess}\ \emph {et~al.}(2001)\citenamefont {Hess},
  \citenamefont {Baumann}, \citenamefont {Ammerahl}, \citenamefont {B\"uchner},
  \citenamefont {Heidrich-Meisner}, \citenamefont {Brenig},\ and\ \citenamefont
  {Revcolevschi}}]{Hess2001}%
  \BibitemOpen
  \bibfield  {author} {\bibinfo {author} {\bibfnamefont {C.}~\bibnamefont
  {Hess}}, \bibinfo {author} {\bibfnamefont {C.}~\bibnamefont {Baumann}},
  \bibinfo {author} {\bibfnamefont {U.}~\bibnamefont {Ammerahl}}, \bibinfo
  {author} {\bibfnamefont {B.}~\bibnamefont {B\"uchner}}, \bibinfo {author}
  {\bibfnamefont {F.}~\bibnamefont {Heidrich-Meisner}}, \bibinfo {author}
  {\bibfnamefont {W.}~\bibnamefont {Brenig}}, \ and\ \bibinfo {author}
  {\bibfnamefont {A.}~\bibnamefont {Revcolevschi}},\ }\href {\doibase
  10.1103/PhysRevB.64.184305} {\bibfield  {journal} {\bibinfo  {journal} {Phys.
  Rev. B}\ }\textbf {\bibinfo {volume} {64}},\ \bibinfo {pages} {184305}
  (\bibinfo {year} {2001})}\BibitemShut {NoStop}%
\bibitem [{\citenamefont {Hess}\ \emph {et~al.}(2007)\citenamefont {Hess},
  \citenamefont {ElHaes}, \citenamefont {Waske}, \citenamefont {B\"uchner},
  \citenamefont {Sekar}, \citenamefont {Krabbes}, \citenamefont
  {Heidrich-Meisner},\ and\ \citenamefont {Brenig}}]{Hess2007a}%
  \BibitemOpen
  \bibfield  {author} {\bibinfo {author} {\bibfnamefont {C.}~\bibnamefont
  {Hess}}, \bibinfo {author} {\bibfnamefont {H.}~\bibnamefont {ElHaes}},
  \bibinfo {author} {\bibfnamefont {A.}~\bibnamefont {Waske}}, \bibinfo
  {author} {\bibfnamefont {B.}~\bibnamefont {B\"uchner}}, \bibinfo {author}
  {\bibfnamefont {C.}~\bibnamefont {Sekar}}, \bibinfo {author} {\bibfnamefont
  {G.}~\bibnamefont {Krabbes}}, \bibinfo {author} {\bibfnamefont
  {F.}~\bibnamefont {Heidrich-Meisner}}, \ and\ \bibinfo {author}
  {\bibfnamefont {W.}~\bibnamefont {Brenig}},\ }\href {\doibase
  10.1103/PhysRevLett.98.027201} {\bibfield  {journal} {\bibinfo  {journal}
  {Phys. Rev. Lett.}\ }\textbf {\bibinfo {volume} {98}},\ \bibinfo {pages}
  {027201} (\bibinfo {year} {2007})}\BibitemShut {NoStop}%
\bibitem [{\citenamefont {Hess}\ \emph {et~al.}(2006)\citenamefont {Hess},
  \citenamefont {Ribeiro}, \citenamefont {B\"uchner}, \citenamefont {ElHaes},
  \citenamefont {Roth}, \citenamefont {Ammerahl},\ and\ \citenamefont
  {Revcolevschi}}]{Hess2006}%
  \BibitemOpen
  \bibfield  {author} {\bibinfo {author} {\bibfnamefont {C.}~\bibnamefont
  {Hess}}, \bibinfo {author} {\bibfnamefont {P.}~\bibnamefont {Ribeiro}},
  \bibinfo {author} {\bibfnamefont {B.}~\bibnamefont {B\"uchner}}, \bibinfo
  {author} {\bibfnamefont {H.}~\bibnamefont {ElHaes}}, \bibinfo {author}
  {\bibfnamefont {G.}~\bibnamefont {Roth}}, \bibinfo {author} {\bibfnamefont
  {U.}~\bibnamefont {Ammerahl}}, \ and\ \bibinfo {author} {\bibfnamefont
  {A.}~\bibnamefont {Revcolevschi}},\ }\href {\doibase
  10.1103/PhysRevB.73.104407} {\bibfield  {journal} {\bibinfo  {journal} {Phys.
  Rev. B}\ }\textbf {\bibinfo {volume} {73}},\ \bibinfo {pages} {104407}
  (\bibinfo {year} {2006})}\BibitemShut {NoStop}%
\bibitem [{\citenamefont {Louis}\ \emph {et~al.}(2006)\citenamefont {Louis},
  \citenamefont {Prelov\ifmmode~\check{s}\else \v{s}\fi{}ek},\ and\
  \citenamefont {Zotos}}]{Louis2006}%
  \BibitemOpen
  \bibfield  {author} {\bibinfo {author} {\bibfnamefont {K.}~\bibnamefont
  {Louis}}, \bibinfo {author} {\bibfnamefont {P.}~\bibnamefont
  {Prelov\ifmmode~\check{s}\else \v{s}\fi{}ek}}, \ and\ \bibinfo {author}
  {\bibfnamefont {X.}~\bibnamefont {Zotos}},\ }\href {\doibase
  10.1103/PhysRevB.74.235118} {\bibfield  {journal} {\bibinfo  {journal} {Phys.
  Rev. B}\ }\textbf {\bibinfo {volume} {74}},\ \bibinfo {pages} {235118}
  (\bibinfo {year} {2006})}\BibitemShut {NoStop}%
\bibitem [{\citenamefont {Kl\"{u}mper}\ and\ \citenamefont
  {Sakai}(2002)}]{Klumper2002}%
  \BibitemOpen
  \bibfield  {author} {\bibinfo {author} {\bibfnamefont {A.}~\bibnamefont
  {Kl\"{u}mper}}\ and\ \bibinfo {author} {\bibfnamefont {K.}~\bibnamefont
  {Sakai}},\ }\href {http://stacks.iop.org/0305-4470/35/i=9/a=307} {\bibfield
  {journal} {\bibinfo  {journal} {Journal of Physics A: Mathematical and
  General}\ }\textbf {\bibinfo {volume} {35}},\ \bibinfo {pages} {2173}
  (\bibinfo {year} {2002})}\BibitemShut {NoStop}%
\bibitem [{\citenamefont {Orignac}\ \emph {et~al.}(2003)\citenamefont
  {Orignac}, \citenamefont {Chitra},\ and\ \citenamefont
  {Citro}}]{Orignac2003}%
  \BibitemOpen
  \bibfield  {author} {\bibinfo {author} {\bibfnamefont {E.}~\bibnamefont
  {Orignac}}, \bibinfo {author} {\bibfnamefont {R.}~\bibnamefont {Chitra}}, \
  and\ \bibinfo {author} {\bibfnamefont {R.}~\bibnamefont {Citro}},\ }\href
  {\doibase 10.1103/PhysRevB.67.134426} {\bibfield  {journal} {\bibinfo
  {journal} {Phys. Rev. B}\ }\textbf {\bibinfo {volume} {67}},\ \bibinfo
  {pages} {134426} (\bibinfo {year} {2003})}\BibitemShut {NoStop}%
\bibitem [{\citenamefont {Rozhkov}\ and\ \citenamefont
  {Chernyshev}(2005)}]{Rozhkov2005}%
  \BibitemOpen
  \bibfield  {author} {\bibinfo {author} {\bibfnamefont {A.~V.}\ \bibnamefont
  {Rozhkov}}\ and\ \bibinfo {author} {\bibfnamefont {A.~L.}\ \bibnamefont
  {Chernyshev}},\ }\href {\doibase 10.1103/PhysRevLett.94.087201} {\bibfield
  {journal} {\bibinfo  {journal} {Phys. Rev. Lett.}\ }\textbf {\bibinfo
  {volume} {94}},\ \bibinfo {pages} {087201} (\bibinfo {year}
  {2005})}\BibitemShut {NoStop}%
\bibitem [{\citenamefont {Saito}(2003)}]{Saito2003}%
  \BibitemOpen
  \bibfield  {author} {\bibinfo {author} {\bibfnamefont {K.}~\bibnamefont
  {Saito}},\ }\href {\doibase 10.1103/PhysRevB.67.064410} {\bibfield  {journal}
  {\bibinfo  {journal} {Phys. Rev. B}\ }\textbf {\bibinfo {volume} {67}},\
  \bibinfo {pages} {064410} (\bibinfo {year} {2003})}\BibitemShut {NoStop}%
\bibitem [{\citenamefont {Saito}\ \emph {et~al.}(1996)\citenamefont {Saito},
  \citenamefont {Takesue},\ and\ \citenamefont {Miyashita}}]{Saito1996}%
  \BibitemOpen
  \bibfield  {author} {\bibinfo {author} {\bibfnamefont {K.}~\bibnamefont
  {Saito}}, \bibinfo {author} {\bibfnamefont {S.}~\bibnamefont {Takesue}}, \
  and\ \bibinfo {author} {\bibfnamefont {S.}~\bibnamefont {Miyashita}},\ }\href
  {\doibase 10.1103/PhysRevE.54.2404} {\bibfield  {journal} {\bibinfo
  {journal} {Phys. Rev. E}\ }\textbf {\bibinfo {volume} {54}},\ \bibinfo
  {pages} {2404} (\bibinfo {year} {1996})}\BibitemShut {NoStop}%
\bibitem [{\citenamefont {Sakai}\ and\ \citenamefont
  {Kl\"{u}mper}(2003)}]{Sakai2003}%
  \BibitemOpen
  \bibfield  {author} {\bibinfo {author} {\bibfnamefont {K.}~\bibnamefont
  {Sakai}}\ and\ \bibinfo {author} {\bibfnamefont {A.}~\bibnamefont
  {Kl\"{u}mper}},\ }\href {http://stacks.iop.org/0305-4470/36/i=46/a=006}
  {\bibfield  {journal} {\bibinfo  {journal} {Journal of Physics A:
  Mathematical and General}\ }\textbf {\bibinfo {volume} {36}},\ \bibinfo
  {pages} {11617} (\bibinfo {year} {2003})}\BibitemShut {NoStop}%
\bibitem [{\citenamefont {Shimshoni}\ \emph {et~al.}(2005)\citenamefont
  {Shimshoni}, \citenamefont {Andrei},\ and\ \citenamefont
  {Rosch}}]{Shimshoni2005}%
  \BibitemOpen
  \bibfield  {author} {\bibinfo {author} {\bibfnamefont {E.}~\bibnamefont
  {Shimshoni}}, \bibinfo {author} {\bibfnamefont {N.}~\bibnamefont {Andrei}}, \
  and\ \bibinfo {author} {\bibfnamefont {A.}~\bibnamefont {Rosch}},\ }\href
  {\doibase 10.1103/PhysRevB.72.059903} {\bibfield  {journal} {\bibinfo
  {journal} {Phys. Rev. B}\ }\textbf {\bibinfo {volume} {72}},\ \bibinfo
  {pages} {059903(E)} (\bibinfo {year} {2005})}\BibitemShut {NoStop}%
\bibitem [{\citenamefont {Shimshoni}\ \emph {et~al.}(2003)\citenamefont
  {Shimshoni}, \citenamefont {Andrei},\ and\ \citenamefont
  {Rosch}}]{Shimshoni2003}%
  \BibitemOpen
  \bibfield  {author} {\bibinfo {author} {\bibfnamefont {E.}~\bibnamefont
  {Shimshoni}}, \bibinfo {author} {\bibfnamefont {N.}~\bibnamefont {Andrei}}, \
  and\ \bibinfo {author} {\bibfnamefont {A.}~\bibnamefont {Rosch}},\ }\href
  {\doibase 10.1103/PhysRevB.68.104401} {\bibfield  {journal} {\bibinfo
  {journal} {Phys. Rev. B}\ }\textbf {\bibinfo {volume} {68}},\ \bibinfo
  {pages} {104401} (\bibinfo {year} {2003})}\BibitemShut {NoStop}%
\bibitem [{\citenamefont {Zotos}\ \emph {et~al.}(1997)\citenamefont {Zotos},
  \citenamefont {Naef},\ and\ \citenamefont {Prelovsek}}]{Zotos1997}%
  \BibitemOpen
  \bibfield  {author} {\bibinfo {author} {\bibfnamefont {X.}~\bibnamefont
  {Zotos}}, \bibinfo {author} {\bibfnamefont {F.}~\bibnamefont {Naef}}, \ and\
  \bibinfo {author} {\bibfnamefont {P.}~\bibnamefont {Prelovsek}},\ }\href
  {\doibase 10.1103/PhysRevB.55.11029} {\bibfield  {journal} {\bibinfo
  {journal} {Phys. Rev. B}\ }\textbf {\bibinfo {volume} {55}},\ \bibinfo
  {pages} {11029} (\bibinfo {year} {1997})}\BibitemShut {NoStop}%
\bibitem [{\citenamefont {Motoyama}\ \emph {et~al.}(1996)\citenamefont
  {Motoyama}, \citenamefont {Eisaki},\ and\ \citenamefont
  {Uchida}}]{Motoyama1996}%
  \BibitemOpen
  \bibfield  {author} {\bibinfo {author} {\bibfnamefont {N.}~\bibnamefont
  {Motoyama}}, \bibinfo {author} {\bibfnamefont {H.}~\bibnamefont {Eisaki}}, \
  and\ \bibinfo {author} {\bibfnamefont {S.}~\bibnamefont {Uchida}},\ }\href
  {\doibase 10.1103/PhysRevLett.76.3212} {\bibfield  {journal} {\bibinfo
  {journal} {Phys. Rev. Lett.}\ }\textbf {\bibinfo {volume} {76}},\ \bibinfo
  {pages} {3212} (\bibinfo {year} {1996})}\BibitemShut {NoStop}%
\bibitem [{\citenamefont {Teske}\ and\ \citenamefont
  {M\"{u}ller-Buschbaum}(1970)}]{Teske1970}%
  \BibitemOpen
  \bibfield  {author} {\bibinfo {author} {\bibfnamefont {C.~L.}\ \bibnamefont
  {Teske}}\ and\ \bibinfo {author} {\bibfnamefont {H.}~\bibnamefont
  {M\"{u}ller-Buschbaum}},\ }\href@noop {} {\bibfield  {journal} {\bibinfo
  {journal} {Z. anorg. allg. Chem.}\ }\textbf {\bibinfo {volume} {379}},\
  \bibinfo {pages} {234} (\bibinfo {year} {1970})}\BibitemShut {NoStop}%
\bibitem [{\citenamefont {Teske}\ and\ \citenamefont
  {M\"{u}ller-Buschbaum}(1969)}]{Teske1969}%
  \BibitemOpen
  \bibfield  {author} {\bibinfo {author} {\bibfnamefont {C.~L.}\ \bibnamefont
  {Teske}}\ and\ \bibinfo {author} {\bibfnamefont {H.}~\bibnamefont
  {M\"{u}ller-Buschbaum}},\ }\href@noop {} {\bibfield  {journal} {\bibinfo
  {journal} {Z. anorg. allg. Chem.}\ }\textbf {\bibinfo {volume} {371}},\
  \bibinfo {pages} {325} (\bibinfo {year} {1969})}\BibitemShut {NoStop}%
\bibitem [{\citenamefont {Hammerath}\ \emph {et~al.}(2011)\citenamefont
  {Hammerath}, \citenamefont {Nishimoto}, \citenamefont {Grafe}, \citenamefont
  {Wolter}, \citenamefont {Kataev}, \citenamefont {Ribeiro}, \citenamefont
  {Hess}, \citenamefont {Drechsler},\ and\ \citenamefont
  {B\"uchner}}]{Hammerath2011}%
  \BibitemOpen
  \bibfield  {author} {\bibinfo {author} {\bibfnamefont {F.}~\bibnamefont
  {Hammerath}}, \bibinfo {author} {\bibfnamefont {S.}~\bibnamefont
  {Nishimoto}}, \bibinfo {author} {\bibfnamefont {H.-J.}\ \bibnamefont
  {Grafe}}, \bibinfo {author} {\bibfnamefont {A.~U.~B.}\ \bibnamefont
  {Wolter}}, \bibinfo {author} {\bibfnamefont {V.}~\bibnamefont {Kataev}},
  \bibinfo {author} {\bibfnamefont {P.}~\bibnamefont {Ribeiro}}, \bibinfo
  {author} {\bibfnamefont {C.}~\bibnamefont {Hess}}, \bibinfo {author}
  {\bibfnamefont {S.-L.}\ \bibnamefont {Drechsler}}, \ and\ \bibinfo {author}
  {\bibfnamefont {B.}~\bibnamefont {B\"uchner}},\ }\href {\doibase
  10.1103/PhysRevLett.107.017203} {\bibfield  {journal} {\bibinfo  {journal}
  {Phys. Rev. Lett.}\ }\textbf {\bibinfo {volume} {107}},\ \bibinfo {pages}
  {017203} (\bibinfo {year} {2011})}\BibitemShut {NoStop}%
\bibitem [{\citenamefont {Callaway}(1959)}]{Callaway1959}%
  \BibitemOpen
  \bibfield  {author} {\bibinfo {author} {\bibfnamefont {J.}~\bibnamefont
  {Callaway}},\ }\href {\doibase 10.1103/PhysRev.113.1046} {\bibfield
  {journal} {\bibinfo  {journal} {Phys. Rev.}\ }\textbf {\bibinfo {volume}
  {113}},\ \bibinfo {pages} {1046} (\bibinfo {year} {1959})}\BibitemShut
  {NoStop}%
\bibitem [{Note1()}]{Note1}%
  \BibitemOpen
  \bibinfo {note} {As the method for extraction of magnetic heat conductivity
  relies on the assumption of isotropic phononic heat conduction, there could
  always be an error in our estimation that stems from the anisotropy of
  phononic conduction parallel and perpendicular to the chains. This error is
  large at low temperatures as the magnitudes of $\kappa _b$ and $\kappa $$_c$
  are large in this regime, thus creating significant uncertainty in the
  extracted $\kappa _{\protect \mathrm {mag}}$. The errors have been calculated
  by taking into account a 30\% anisotropy in the phononic heat conductivity
  parallel and perpendicular to the chains.}\BibitemShut {Stop}%
\bibitem [{\citenamefont {Hess}(2007)}]{Hess2007}%
  \BibitemOpen
  \bibfield  {author} {\bibinfo {author} {\bibfnamefont {C.}~\bibnamefont
  {Hess}},\ }\href {\doibase 10.1140/epjst/e2007-00363-8} {\bibfield  {journal}
  {\bibinfo  {journal} {The European Physical Journal Special Topics}\ }\textbf
  {\bibinfo {volume} {151}},\ \bibinfo {pages} {73} (\bibinfo {year}
  {2007})}\BibitemShut {NoStop}%
\end{thebibliography}%

\end{document}